\documentclass[prl, twocolumn]{revtex4-2}
\usepackage{amssymb}
\usepackage{amsmath}
\usepackage{epsfig}
\usepackage{bm}

\begin{document}

\title{ Non-Newtonian two-dimensional electron fluid in magnetic field }

\author{  P. S. Alekseev and  M. A. Semina }

\affiliation{  Ioffe  Institute,  194021  St.~Petersburg, Russia }

\begin{abstract}

We develop a theory of the non-Newtonian regime of non-linear hydrodynamic  magnetotransport of
a two-dimensional (2D) viscous electron fluid controlled by the local Joule heating.
In this mechanism, the electron shear viscosity in magnetic field becomes
a non-monotonic function of the gradient of the hydrodynamic velocity
due to the flow-induced increase   of the electron temperature.
We derive  and solve the corresponding non-linear hydrodynamic equations
 for a velocity profile in a Poiseuille-like flow geometry.
We demonstrate that the  calculated magnetoresistance of such flow
well explains  the differential magnetoresistance observed for 2D electrons
 in various samples of ultra-high-quality GaAs quantum wells at high currents.  We conclude that
the  non-linear regime of 2D electron fluid flows is such  systems   is realized
via formation of a 2D non-Newtonian electron fluid, related to non-linearity in viscosity,
but not via convective ``kinematic-induced'' effects, as it takes place in ordinary uncharged fluids.

\end{abstract}

\maketitle

{\em 1. Introduction.} Hydrodynamic electron transport was realized in recent ten years
 in conductors with very low densities of defect at the condition of frequent
interelectron collisions. It was first reliably identified
 in ultrapure samples of layered palladium cobaltate~\cite{Moll},
in single-layered graphene~\cite{graphene_1,Polini_Geim,Levitov_et_al},
 and in gallium arsenide quantum wells. In the latter case,
 this regime  was first identified by the strong decrease
 in the sample resistance with the increase of magnetic field
(the giant negative magnetoresistance), induced
 by the magnetic field dependence of the electron  viscosity  \cite{Gusev_1,je_visc,recentest_,exps_neg_1,exps_neg_2,exps_neg_3,exps_neg_4,Gurzhi_Shevchenko},
and, then, by the dependence of the sample  resistance on its complex
geometry~\cite{Keser}.

In recent ten years many bright effects of the hydrodynamic 2D electron transport
 were studied experimentally and theoretically.
Essentially two-dimensional flows of the electron fluid   in samples with
 macroscopic obstacles  were examined~\cite{d1,d1_new,d2,d2_new,disks,disks2}.
In Refs.~\cite{Scaffidi2017,a,Holder,c}
 the effects of the transition from a hydrodynamic to a ballistic
 flow regime with an increase in the magnetic field
 were investigated for very pure samples.
 Other stationary and low-frequency hydrodynamic transport effects
were considered in many works, in particular, in
 Refs.~\cite{Afanasiev2022,Denisov2022,Glazov,Denisov2023,Alekseev_2023,
 Afanasiev_at_al_2025,Alekseev_Dmitriev,Polini,Levin2024}.
 Unexpected effects of high-frequency hydrodynamic transport
  were theoretically studied
in Refs.~\cite{vis_res_0,vis_res_1,vis_res_2,Semiconductors,Afanasiev_2023,new},
and, possibly, were observed  in best-quality samples
in experiments~\cite{Smet,Dai2010,Hatke2011,Bandurin2022}
(see a detailed discussion  in~\cite{new}).

Non-linear effects  in 2D electron hydrodynamics also attracts
  a great attention. From the one hand,
 manifestation of the non-linear effects, induced by  the convective mechanism
due to the kinematic term $( \mathbf{V} \cdot \nabla) \mathbf{V}  $,
is typically associated with requirements for  exotic  channel and contact  designs
or flow regimes that are difficult to  achieve experimentally~\cite{1n}.
  Realization of such effects is limited, for example, by thermal dissipation and
the presence of residual impurities~\cite{3n,4n}.
  Characteristic Reynolds numbers corresponding to record-high electron flow velocities
lie in the range from~$\mathrm{Re}=10^{-2}$ to~$\mathrm{Re}=1 $~\cite{1n},
far from the threshold  of the  formation of
 essentially non-linear waves (such as considered in well-known work~\cite{Dyakonov_Shur})
or unsteady  turbulent flows,~$\mathrm{Re} \sim 10^3$.
Nevertheless several striking experimental result
 on non-linear effects in the electron fluid  were achieved recently.
Implementation in bilayer-graphene-based structures of the Laval electron nozzle,
which generates supersonic outflow  of the electron fluid, was demonstrated
in Ref.~\cite{2n}.   In Ref.~\cite{8n} an electron analogue of the Tesla valve,
a passive fluidic diode   that rectifies a flow, was realized
in   high-quality GaAs quantum well structures.

On the other hand, recent research increasingly  highlights the importance of other type
non-linear effects in electron hydrodynamics, namely rheological effects,
 that is, the non-linearities associated with the dependence of the fluid viscosity
 on the flow velocity profile~\cite{rheol,rheol2}.
 A number of microscopic  mechanisms have been proposed
that may be responsible for such non-Newtonian behavior of 2D electron fluids:
(i)~the peculiarities of the relaxation of even and odd angular harmonics
 of the nonequilibrium distribution function (the so-called tomographic
regime)~\cite{5n}; (ii)~broken symmetry with respect to the inversion
 of time and space in  the material of a sample~\cite{6n,6n2};
(iii)~the memory effects in  electron-electron collisions in
classically strong magnetic fields~\cite{Alekseev_Semina_2025}.

In parallel with the above mentioned works,
 experimental studies of the  non-linear dc regime of
2D electron magnetotransport in ultra-pure samples of GaAs quantum wells were
 performed~\cite{non-lin_hydr_1,non-lin_hydr_3,non-lin_hydr_4,non-lin_hydr_5,non-lin_hydr_6}.
In particular, it was shown that the giant negative magnetoresistance,
 being characteristic for stationary flows of a viscous electron fluid,
  changes with the increase of the current in a characteristic way:
it becomes strongly non-monotonic in the region of small fields;
the resistance decreases with the current near the zero magnetic field,
and grows  with the current at the large
  fields~\cite{non-lin_hydr_1,non-lin_hydr_5,non-lin_hydr_6}.
Up to now there is no consistent microscopic theory which fully explains
this effect within some non-linear hydrodynamic model.

In this Letter we develop a microscopic theory of the non-Newtonian regime
 of a 2D electron fluid based on a straightforward mechanism
 for the rheological  non-linearity, being the local Joule heating,
that is the local increase of the electron temperature
 due to  heat dissipation and the corresponding change of viscosity.
The electron temperature in each sample region is controlled by the balance
between the heat production  in a viscous flow and the heat exchange
with acoustic phonons in the structure.  We derive the non-linear
Navier-Stokes-like equation and calculate the  distribution
of the hydrodynamic velocity  for a Poiseuille-like flow
and find the corresponding  differential magnetoresistance.
 The last one explains very well the nonlinear differential
magnetoresistance observed   on several different samples examined in independent
  experiments~\cite{non-lin_hydr_1,non-lin_hydr_5,non-lin_hydr_6} on high-quality
GaAs quantum wells. This result evidences that  the  non-linear regime of
2D electron fluid flows in such systems is typically realized
via formation of a 2D non-Newtonian electron fluid,
but not via convective ``kinematic-induced'' effects~\cite{f1}.

{\em 2. Model for non-linear magnetotransport of 2D electron fluid. }
 We consider the flows of 2D degenerate electrons
 for  the simplest flat geometry in long samples,
the  Poiseuille-like flows $V_x=V(y,t)$ [see Fig.1(a)].

In Refs.~\cite{el,el2,el3}  hydrodynamic-like viscoelastic
 models of ac flows of a
2D electron fluid, based on the Fermi-liquid theory, were developed.
Within a similar approach, in Refs.~\cite{vis_res_2,Semiconductors}
  a  model of flows of a highly-correlated viscous
 2D electron fluid  in classical magnetic fields was constructed.
   In Ref.~\cite{new,Alekseev_Semina_2025},
   a phenomenological theory of the non-linear electron magnetotransport,
 accounting the pair electron-electron  correlations
within a classical-mechanics picture (that is, the memory effects
in the inter-particle scattering), was constructed.
The pair correlations  are  induced by subsequent (``extended'') collisions
of some electrons joined   in pairs due to the cyclotron rotation
\{other  electrons do not
    do not scatter the cyclotron period
   or scatter without memory effects;
   see  Fig.~1(a) and Refs.~\cite{new,Alekseev_Semina_2025}\}.
 The pair correlations lead to the non-linear retarded  terms
 in the Navier-Stokes-like hydrodynamic equations~\cite{new,Alekseev_Semina_2025}.

\begin{figure}[t!]
\centerline{\includegraphics[width=.99 \linewidth]{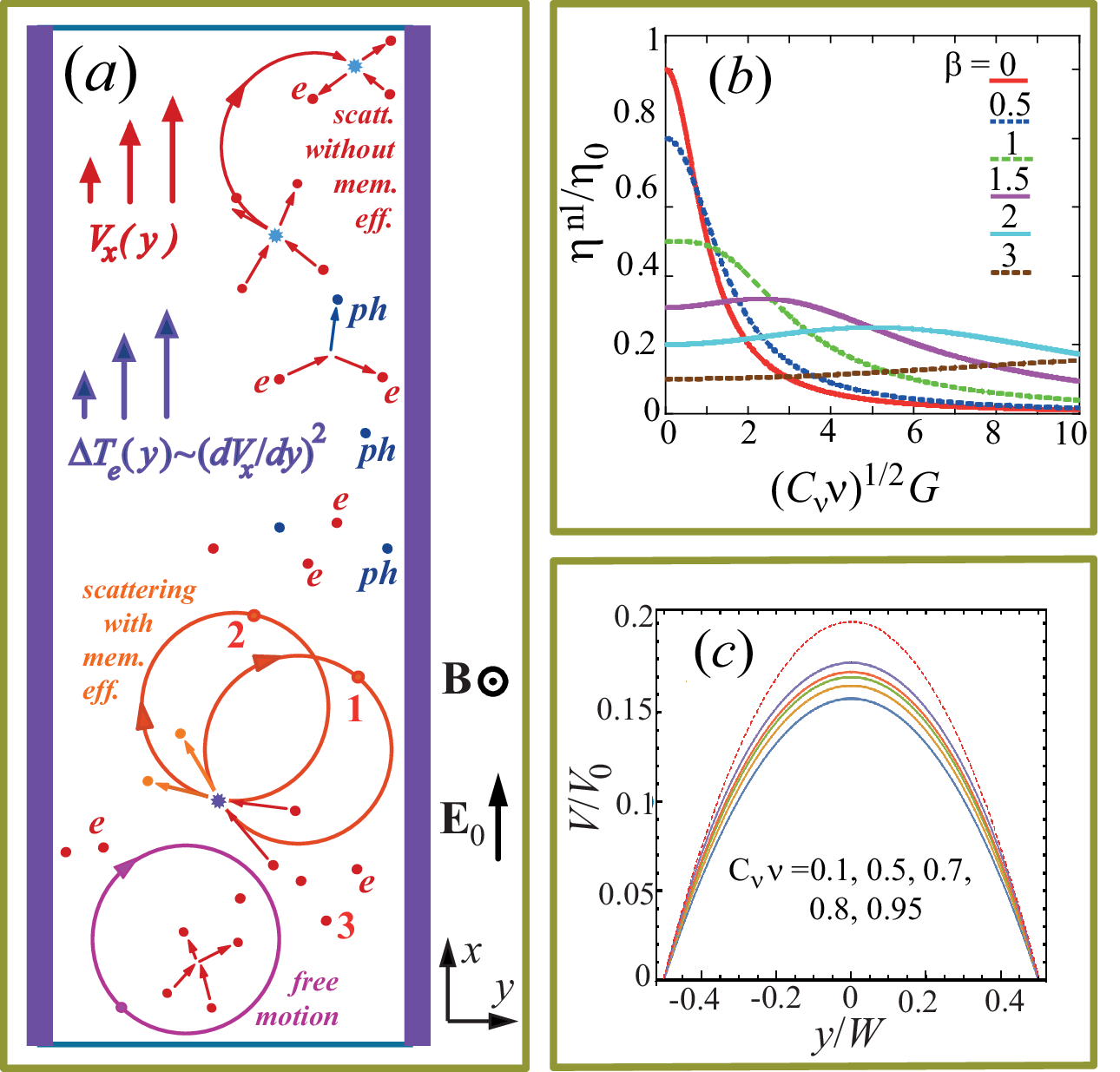}}
\caption{
($a$):
Long sample with 2D electrons in magnetic and electric  fields;
 classical-mechanics sketches
 of electron-electron and electron-phonon scattering events
and the inhomogeneous electron temperature profile.
($b$):
Electron viscosity accounting the local heating effect as function of
 the dimensionless nonlinearity parameter
  $ \sqrt{ C_\nu \nu } \,   G $
 at different magnetic fields
($G=dV_x/dy$).
 It is seen that with the increase of
magnetic field the function
 $ \eta^{nl} (  G)$ becomes non-monotonic
($c$):
The velocity profile  of the electron fluid flow calculated within
 the proposed mechanism of non-linearity for the parameter
   $\beta=0.5 $ and increasing values of~$  C_\nu \nu $ for
   the curves from the blue  one  to the violet  one.
 The dashed red line is the parabolic profile  with the same derivatives
  at the edges as for the violet curve,
  shown for demonstration of the non-parabolic shape
    of the violet curve.
  }
\end{figure}

There also exists a simple mechanism of non-linearity related
 to the local heating of electrons.
When a viscous flow of any fluid is formed,
 a Joule heat (that is, the heat due to dissipation processes)
is released, leading
 to non-linear properties of the flow. In this work we consider
this mechanism separately from others mechanisms of non-linearity.

 Density of the heat release in a viscous flow per one electron
  is $\dot{Q}_v = m \eta_{xx} (\partial V / \partial y  )^2$, where $m$
 is the electron effective mass;
 $\eta_{xx} =\eta_0/(1+\beta^2) $ is the kinematic diagonal viscosity,
$\beta = 2\omega_c \tau_{2,ee}^{(0)}$, $\eta_0 = v_F^2 \tau_{2,ee}^{(0)}/4 $, and
  $\tau_{2,ee}^{(0)} = \tau_{2,ee}(T=T_{ph})$ is the shear stress relaxation time
  due to inter-particle collisions at the lattice temperature $T_{ph}$.
If we neglect thermoconductivity and thermoelectric effects,
this heat release is compensated by the heat exchange
 with acoustic phonons. The rate  of the last processes at $T_e \sim  T_{ph}$
 is equal to   $ \dot{Q}_{tr} = \alpha \,(T_e-T_{ph})$, where $T_e $
 is the electron temperature, which depends of the coordinate $y$ in a flow
and the coefficient  $\alpha = C_\alpha m E_D^2 / (T_{ph} \varrho \hbar a^3 )$
was calculated for GaAs  quantum well structures in Ref.~\cite{Karpus_1}.
 In it, $E_D$ is  the deformation potential for the electron interaction
with acoustic phonons, $\varrho$ is the crystal density, $s$ is the speed of sound,
  $a$ is the width of the quantum well, and $C_\alpha$ is a numeric constant
of the order of unity depending of the shape of the quantum well potential.
In this way, we obtain the closed equation for the electron temperature $T_e(y)$:
\begin{equation}
    \label{T_e}
  \alpha \,  [ \, T_e(y) - T_{ph} \,  ]
   =  m \,  \eta_{xx}  \, ( \,  \partial V / \partial y \, ) ^2
 \:.
\end{equation}

The rate  of the relaxation of the shear stress is given by the formula:
$\hbar / \tau_{2,ee} = C_{\tau} T_e^2 /
 \varepsilon _F$~\cite{Novikov,Polini,Alekseev_Dmitriev},
   where $C_\tau$ is a smooth function of temperature and
 the interaction parameter $r_s = 1/(\sqrt{\pi n_0} \, a_B)$,
  having the values of order of unity in realistic 2D Fermi systems. Here
$n_0$ is the 2D electron density and  $a_B$ is the Bohr radius.
 Provided the difference $\Delta T_e(y) = T_e(y) - T_{ph} $ is still smaller that $T_{ph}$,
 we obtain the expression for the
  increase of the shear stress relaxation rate
  due to the local electron heating:
\begin{equation}
    \label{tau_y}
     \frac{1}{\tau_{2,ee}(y)} =
      \frac{1}{\tau_{2,ee}^{(0)}}
+ \tau_\Delta \Big( \frac{\partial V }{\partial y }  \Big ) ^2 ,
\end{equation}
where  $\tau_\Delta = [C_{\alpha} \varrho \hbar a^3 v_F^2 /(2E_D^2)]/(1+\beta^2)$.

Here we consider the simplest
 ``quasi-one-dimensional''  flows
    in long and relatively  narrow samples: with
the width $W$ much smaller  than the plasmon wavelength \{see Fig.~1(a)
and Refs.~\cite{vis_res_2,new}\}.
Here
the $y$-component of  the flow velocity, $V_y$, related
to an internal electric  field $\mathbf{E}_y ^{int} (y,t) $ and
a non-equilibrium charge density~$ e \,\delta n (y,t)$,
is suppressed~\cite{vis_res_2}.  As a result, the motion equations
for the hydrodynamic velocity $ V(y,t)  \equiv V_x (y,t) $ and
 the shear stress~$  \hat{\sigma} (y,t)  = -\hat{\Pi} (y,t) $
take the form~\cite{f2}:
  \begin{equation}
\label{main_eq_gen}
\left\{
\begin{array}{l}
\displaystyle
    m \: \frac{   \partial V   }{ \partial t   }  =
          e   E   -  \frac{  \partial \Pi_{xy}  }{ \partial y}
 \:,
\\
\displaystyle
 \frac{\partial \Pi_{xx} }{\partial  t } \, = \,  2 \omega_c  \Pi_{xy}
   \, -\,  \frac{ \Pi_{xx} }{ \tau_{2}(y)}
          \: ,
\\
\displaystyle
 \frac{\partial \Pi_{xy} }{\partial  t }
 = - \frac{ \Pi_{xy} }{ \tau_{2}(y) }  - 2 \omega_c  \Pi_{xx}  -
    \frac{   m \eta_0 }{ \tau_{2,ee}^{(0)} }
       \: \frac{  \partial V }{ \partial{y} }  \, .
\end{array}
\right.
\end{equation}
where  $ 1/\tau_2 (y) = 1/\tau_2 [ \partial V /\partial y]  $
is the shear stress relaxation rate~(\ref{tau_y}) due to the local heating;
 the slow electric field
$E \equiv E_x (t) $ is due  to the applied voltage.

For the slow varying flows, with the frequencies $\omega$
much smaller that the shear stress relaxation
rate~$1/\tau_2(y) \sim 1/\tau_{2,ee}^{(0)}$.
   we omit the time derivatives  of $\hat{\Pi}$
    in~(\ref{main_eq_gen}), and obtain
that the second and the third lines in Eq.~(\ref{main_eq_gen}) yield
 to the following  relation between $\Pi_{xy}(y,t)$
and~$\partial V (y,t)/ \partial y  $:  $  \:   \Pi_{xy}  \,
=\,    - \, m\, \eta _{xx}^{nl}  [V] \: \partial V/ \partial y  \: , $
where we introduce the  non-linear viscosity
coefficient~$ \eta _{xx}^{nl}  [V] (y,t)$:
\begin{equation}
\label{eta_nl}
 \begin{array}{c}
  \displaystyle
 \eta _{xx}^{nl}  [V]  = \eta_0 \, / \, ( \,
   h + \beta  ^2   /  h \,  )
   \:,
   \\\\
   \displaystyle
\beta  =  2 \omega_c \tau_{2,ee}^{(0)}
\:,
\quad
   h = 1 + \tau_{2,ee}^{(0)}  \, \tau _\Delta \,
    ( \, \partial V  /   \partial y \, ) ^2
   \,.
   \end{array}
\end{equation}
The resulting motion  equation for such slow flows  is:
\begin{equation}
\label{motion_eq_fin}
   \frac{ \partial V }{ \partial t }
    \, = \,
   \frac{eE}{m}
   \,   + \,
   \frac{\partial}{\partial y} \Big( \, \eta _{xx}^{nl}  [V] \,
     \frac{ \partial V}{ \partial y } \, \Big)
     \, .
\end{equation}
Formula (\ref{eta_nl}) establishes  that
the local heating leads to the dependence
 of viscosity on the velocity gradient
 $ G=\partial V/\partial  y $. In other words,
 the electron fluid becomes non-Newtonian, namely, pseudo-plastic
at classically low magnetic fields
  (that is, $\eta_{xx}^{nl} $  decreases with
$ G $ at $ \beta   < 1 $) and dilatant   at high magnetic fields
and at relatively small $G$  (that is, $\eta_{xx}^{nl}$ increases
with $G $ and  $ \beta   > 1 $),  while being also pseudo-plastic
 at relatively enough large $G$ again
[see Fig.1($b$)].   Note that
 a non-monotonous behavior of  $\eta_{xx}^{nl} (G)$  with $G$
is atypical  for uncharged non-Newtonian fluids~\cite{rheol2}.

{\em 3. Stationary flow in long sample. }
Now we consider a stationary Poiseuille flow   in a defectless long sample
as a minimal model to study magnetotransport in the non-Newtonian 2D electron
fluid in a non-linear regime.

Let us introduce the dimensionless variables $\widetilde{y} = y/W$
and $\widetilde{V} = V/ V_0$,
where $V_0 = eE W^2 / (m\eta_0) $ is
the characteristic magnitude of the hydrodynamic velocity
in the stationary Poiseuille flow   in the linear by $E $~regime
in the absence of magnetic field.  The symmetry of~$V(y)$
 with respect to the sample center, $y=0$,
  leads to the following form of equation~(\ref{motion_eq_fin})
   in the introduced variables:
\begin{equation}
\label{motion_eq_stat}
   \frac{ \displaystyle  d \widetilde{V} / d \widetilde{y} }
   { \displaystyle
  h +
   \beta ^2/ h }
   =
   - \widetilde{y}
   \, ,
   \quad \;\;
    h =   1 + \nu' \,   \Big( \frac{d \widetilde{V} }{ d\widetilde{y}}  \Big) ^2
  \, ,
\end{equation}
where  $\nu' (\beta)  \,= \,  C_\nu \, \nu \, f(\beta)  $
is the dimensionless parameter of non-linearity,
\begin{equation}
\label{f_,_nu}
  f(\beta)  =  \frac{ 1 }{ 1+ \beta^2}
 \, , \quad \;\;
  \nu  =   \frac{ \varrho \, \hbar \, a^3 v_F^2 }{E_D^2 \, \tau_{2,ee}^{(0)} }
    \, \Big( \, \frac{ eEW }{ m v_F^2 } \, \Big)^2
 \,,
\end{equation}
 $ \nu    $ is the magnetic-field-independent parameter  of non-linearity,
 and~$ C_\nu $ is a numeric coefficient related
 to  $ C_\alpha $ in the above formula for $ \alpha $ as:
$ C_\nu =C_\alpha 4^2/2 = 8 C_\alpha $.

One should impose some boundary condition on the velocity $V(y)$
at the sample edges,
$y=\pm W/2$. For simplicity, we chose the sticking
boundary conditions corresponding
to the very rough edges: $ V|_{y=\pm W/2} = 0 $.

Equation~(\ref{motion_eq_stat}) is the ordinary differential equation,
not containing the unknown function $\widetilde{V}(\widetilde{y})$, but containing
 only $ d\widetilde{V} / d\widetilde{y} $ and  $\widetilde{y}$.
It is solved analytically  and the solution is expressed
 in the parametric form:
\begin{equation}
 \label{solution}
  \begin{array}{l}
  \displaystyle
  \widetilde{y}(u) = - \frac{ (1+ \nu'u^2) \, u  }{ \beta^2 + (1 + \nu'u^2)^2 }
  \:,
  \:\;\;
  \widetilde{V} (u) = \widetilde{ V} _1 - \frac{Y(u)  }{\nu'}\:,
  \\
  \\
  \displaystyle
Y(u)  =
    \frac{ \beta^2 +  1+ \nu'u^2   }{ \beta^2 + (1+ \nu'u^2) ^2 }
    +
    \frac{\ln[  \beta^2 + (1 + \nu'u^2) ^2  ] }{4  }
    \,.
 \end{array}
\end{equation}
where $u $ is the parameter which varies
in the interval $ (-u_0,u_0) $, whose edges corresponds
to the sample edges: $ \widetilde{y}( \pm u_0) = \pm  1/2 $,
  and  $ \widetilde{V}_1 $
 is the constant that allows to satisfy the boundary
conditions~$\widetilde{V}( \pm u_0 ) =0$.   The equation
for the boundary points $\pm u_0$,
$ \widetilde{y}( u_0) = 1/2 $, is the fourth order algebraic equation,
 its proper is real and minimal by the absolute value.
 Then the interval $(-u_0, u_0)$  is mapped  one-to-one
by the function $ \widetilde{y}(u)$  on the interval $(-1/2,1/2)$ and
the function $\widetilde{V} (u(\widetilde{y}))$ is well defined.

\begin{figure}[t!]
\centerline{\includegraphics[width=0.9 \linewidth]{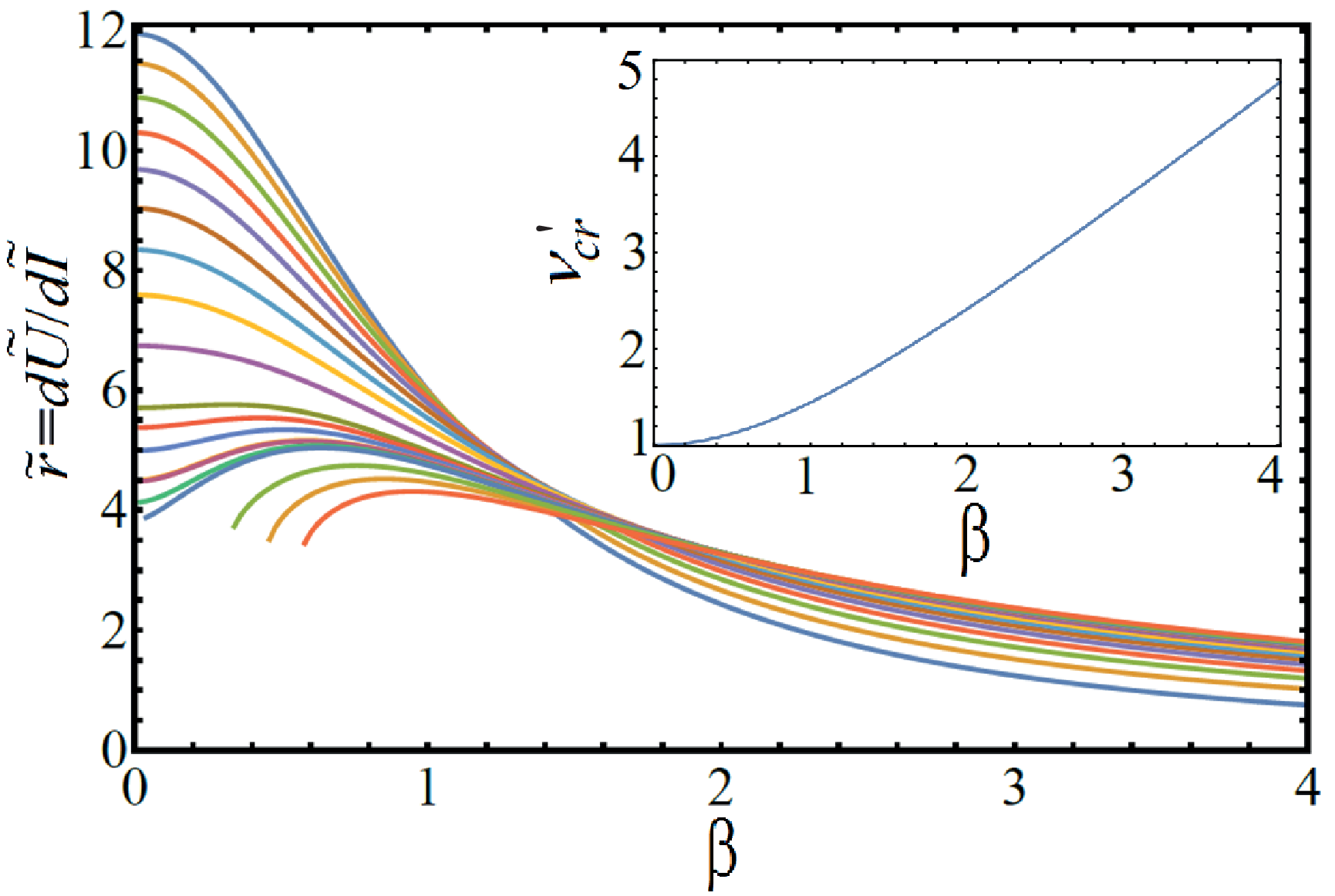}}
\caption{
Calculated dimensionless
differential magnetoresistance
$\tilde{r} = d\tilde{U}/d\tilde{I}$ for
the magnetic-field-independent non-linearity parameters
$C_{\nu} \nu$ = 0.01, 0.1, 0.2, 0.3, 0.4, 0.5,
0.6, 0.7, 0.8, 0.9, 0.925, 0.95, 0.975, 0.98, 0.99,
1.05, 1.1, 1.15.
Inset:
Critical values of
the non-linearity parameters
 $ \nu'_{cr}(\beta) = C_\nu \nu_{cr}(\beta) f(\beta) $.
  }
\end{figure}

The flow profile calculated  by Eq.~(\ref{solution})
is shown in Fig.~1(c). It is seen
 that the dependence of the viscosity
on the velocity gradient  leads to not too large deviation
of the profile   shape from the parabolic
 one, corresponding to
 the linear Poiseuille flow: the flow profile
of the non-Newtonian fluid becomes ``more blunt''.

The electron temperature profile corresponding to Eqs.~(\ref{T_e})
 and~(\ref{solution}) is given by:
\begin{equation}
    T_e(y)  \, = \,  T_{ph}
   \, + \,
    \big[\, m \,  \eta_{xx} V_0 / (\alpha \,  W) \, \big]
   \,
    \big[  \, \widetilde{ V} ' (u)   /    \widetilde{ y} \, ' (u)  \, \big] ^2
  \:.
\end{equation}
According to Fig.~1($c$), this profile  $T_e(y)$  is close to the one
for a linear Poiseuille flow, $V(y)\sim y^2$.

{\em 4. Results and  discussion.  } By use of a numerical analysis of
the obtained  analytical solution~(\ref{solution}),
we have determined  the critical values of
the non-linearity parameters  $ \nu'_{cr}(\beta) $ and $ \nu _{cr}(\beta)
= \nu'_{cr}(\beta)  /[C_\nu  f(\beta) ]$.   Namely,
 below  $ \nu '_{cr} (\beta) $,  $ \nu ' < \nu '_{cr} (\beta) $,
 the functions $u(y)$  and $V(y)$ remails well-defined (unambiguous)
 and our  model leads to a smooth flow profile $V(y)$.
  The function $ \nu '_{cr} (\beta) $ is larger the unity, $ \nu ' =1 $ at any $\beta$,
  thus the proper solution $V(y)$ at given $C_\nu \nu  >1 $ exists
in the diapason of $\beta$, $\beta > \beta_{cr}(\nu)$
 (see Fig.~2).  Herewith the function $\beta_{cr}(\nu)$ grows
  monotonically with $\nu$. Note also that as $f(\beta) \leq 1$,
then at given $C_\nu \nu <1$ the proper solution exist at any~$\beta$.

\begin{figure}[t!]
\centerline{\includegraphics[width=1.09 \linewidth]{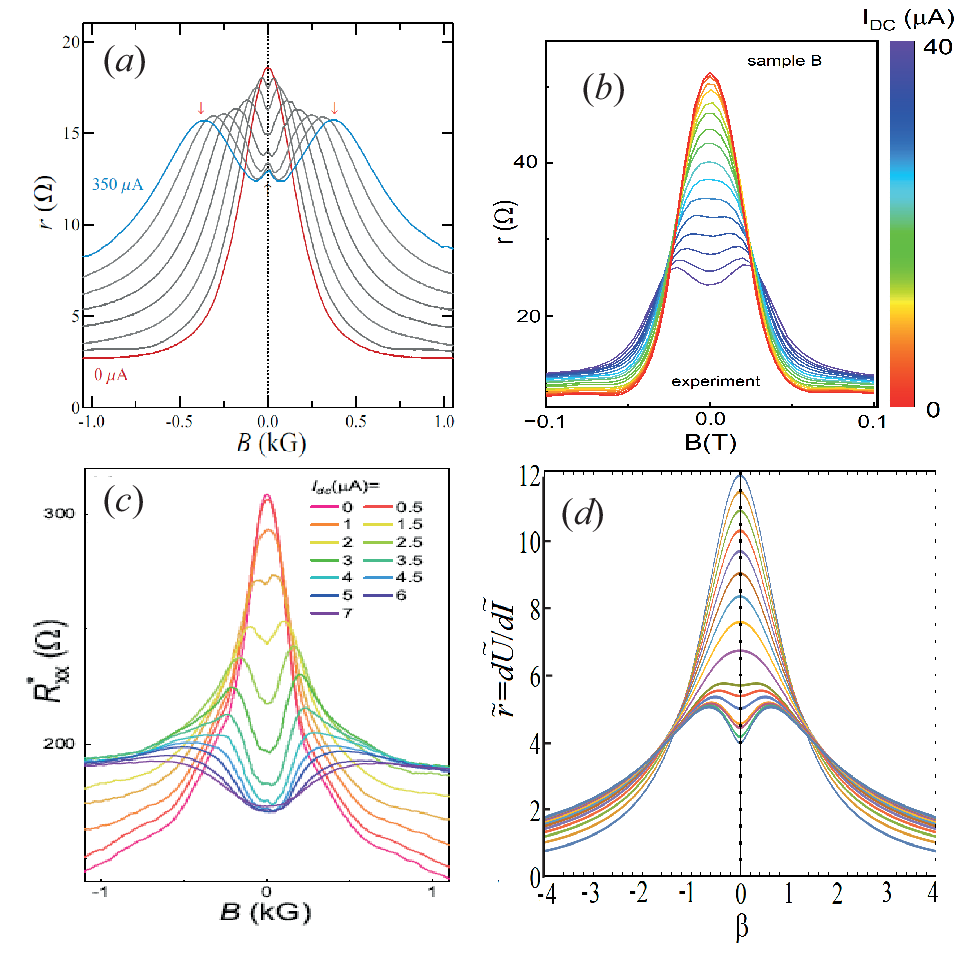}}
\caption{
($a$):
Experimental data on the differential magnetoresistance $r=dU/dI$ of
a high-quality sample of GaAs quantum well at different   values
of the current magnitude obtained in Ref.~\cite{non-lin_hydr_1}.
($b$): The same as in panel (a) from Ref.~\cite{non-lin_hydr_5}.
($c$): The same as in panel (a) from Ref.~\cite{non-lin_hydr_6}.
($d$):
The results of our calculation of the dimensionless
differential magnetoresistance
$\tilde{r} = d\tilde{U}/d\tilde{I}$ for
the magnetic-field-independent non-linearity parameters
$C_{\nu} \nu$ = 0.01, 0.1, 0.2, 0.3, 0.4, 0.5,
0.6, 0.7, 0.8, 0.9, 0.925, 0.95, 0.975, 0.98, 0.99
[the increase of~$C_{\nu} \nu$  corresponds to the decrease
 of the values~$\tilde{r} (\beta =0)$].
  }
\end{figure}

 At higher values of   the non-linearity parameter, $ \nu(\beta) >\nu _{cr}(\beta)   $,
  the function $\widetilde{y}(u)$ in Eq.~(\ref{solution})
becomes non-monotonic, that can correspond to unstable space-inhomogeneous
 solutions.  The last ones can be found only
within some more general model, accounting additional effects
 those stabilize the flow shape  (the effects of
the sample edges, other type of relaxation processes,
in particular, relaxation of the electron momentum
on the residual defects in the bulk of the sample).

The total electric current  can be presented as:  $I=I_0 \, \widetilde{I}$,
where  $ I_0 = e^2 n_0 E W^3 / (m \eta_0 ) $ is its amplitude
in the linear regime, and $\widetilde{I} = \int_{-1/2} ^{1/2} d \widetilde{y}
\: \widetilde{V}( \widetilde{y} )$.  It is of interest
to find the differential resistance  $r=dE/dI = 1/(dI/dE) $, $r=r(\beta,\nu)$,
 being the value which exhibits the features
of  the dependence $I(B,E)$ more clearly.  From the form of the factor~$ I_0$
 and  the nonlinearity parameter~$\nu (E) $, we obtain:
$  r (E)^{-1}\, = \, dI/dE \,  =\, (I_0/E) \, (\,\widetilde{I}    \,
+\ 2\nu \, d \widetilde{I} / d \nu \, )    $.
Using this formula and Eqs.~(\ref{solution}), we calculate the current $I$,
 the resistance $\varrho = E/I$, and the differential
 resistance  $r =  (E/I_0) \,  \tilde{r} $ as functions
of the magnetic field parameter~$\beta$ at different values of~$\nu$.

We have estimated the dimensional non-linearity parameter
$\nu$ by Eq.~(\ref{f_,_nu})
 for characteristic parameters of the sample and
  the experimental conditions
 studied in work~\cite{non-lin_hydr_1}
(for the estimates of the shape and  size of the electron flows and
of the relaxation parameters in this sample see
 Ref.~\cite{je_visc}). We have obtained the value $\nu \sim 0.3$
at the maximal current [see Fig.~3(a)] and the estimated value
for the effective width  of the quantum well  $a \sim 10^{-6}$~cm.
It is a reasonable result that justifies  the applicability of
the local heating mechanism   for the description
of non-linear regime studied in Ref.~\cite{non-lin_hydr_1}.

Fig.~3 demonstrate a good  qualitative agreement
 between all the three sets of
 experimental data on the differential
magnetoresistance of GaAs quantum wells
 and our theoretical results.
One can see the change of the monotonous
 Lorenzian shape of $r(B) = \varrho_{xx} (B)  $
 at $I \to 0 $ ($\nu \to 0$)
  on the non-monotonous shape of the curves $r(B)$ with
  the increase of $I $ and $\nu $.
   A local minimum of $r(B)$  at $B=0$ appears
    and therefore   local maximums are formed
   at some characteristic fields $\pm B_0(T)$,
   in the region where $\beta \sim 1 $.
   Although some particular features
   of the experimental curves differs
   in these three experiments, their general behaviours
    including several characteristic properties
     are  very similar one's to others' as well as to
    the results of our theory [Fig.~3($d$)].
For example, we see from Fig.~3 that the distinct dependencies
 of $r$ on $I$ at $B  \to 0 $ [the decrease of $r(I)$]
and at $B \gg B_0$ [the increase of $r(I)$], are observed in all
experiments~\cite{non-lin_hydr_1,non-lin_hydr_5,non-lin_hydr_6}
and is well explained by our theory.

It is possible that other effects, such as thermal conductivity,
 the thermoelectric effect, and other nonlinearity mechanisms,
for example, the pair electron correlations studied in Ref.~\cite{Alekseev_Semina_2025},
may also be important for the nonlinear  magnetotransport
in particular samples.
 However, the agreement obtained in this study between the predictions
 of our theory of nonlinearity based on the local heating effect
 and the experimental data, in our opinion, allows us to conclude that
  the dependence of the electron viscosity on the flow gradient
(that is, the non-Newtonian behavior in the electron fluid)
plays a key role  for nonlinear hydrodynamic regimes.

{\em 5. Conclusion. } We have theoretically studied the non-Newtonian
behavior of a 2D electron fluid due to local heating.
 We have shown that the type of dependence of nonlinear viscosity
on the velocity gradient crucially depends  on the flow parameters.
We have considered the simplest Poiseuille-like   flow geometry,
 but our model opens the possibility of considering a variety of nonlinear flows
 with more complex 2D geometries. We have demonstrated that the dependencies
of the differential resistance on magnetic field, calculated within our model,
 well explain  the observed nonlinear magnetoresistance of   2D electrons
in several different ultra-pure GaAs quantum wells samples.

 Based  on this result,
we conclude that non-Newtonian electron fluid  is implemented in these 
 (and, possibly, in other similar graphene-based) hydrodynamic conductors
in the non-linear regime by the current magnitude.
An actual  understanding of the mechanism of non-linear electron hydrodynamics 
 in nanostructures opens new possibilities of characterization of these systems 
 as well as developing electronic devices, for example, 
 bistable elements, based on electron flows in strongly non-linear regimes.

We thank M. I. Dyakonov for attracting
our attention to the experimental  results
 on non-linear magnetotransport (first of all, results
 of Ref.~\cite{non-lin_hydr_1}),  on explanation
  of which is aimed this work,
 as well as for fruitful discussions and kind support.
We thank A. N. Afanasiev for attracting our attention
 to the concepts
of non-Newtonian fluids, those lead to the current view
on our results. We thank G. M. Gusev
 for fruitful discussions
 as well as for his interest, comments, and remarks.

Part of this work (the analytical calculations,
the derivation of the nonlinear hydrodynamic equation and
the analytical solution; the model for nonlinear magnetotransport
of the 2D electron fluid, in particular, model for stationary non-linear flow in
long samples) was financially supported by the Russian
Science Foundation (Grant No. 25-12-00093).
Part of this work (the
numerical calculations, the numerical solution of the nonlinear
hydrodynamic equation; the comparison of the obtained
theoretical results and the experiment) was carried
  out within the state assignment of
the Ministry of Science and Higher Education of the Russian
Federation.


\begin{thebibliography}{}
\bibitem{f2}
We note that motion equation~(\ref{main_eq_gen}) does not account other inhomogeneous and non-linear effects,
such as the convection term $(\mathbf{V} \cdot \nabla )\mathbf{V}$;
 the thermoelectric effects; and the enegry transfer effects (see Refs.~\cite{Volkov,Semina,Yassievich}).
\\
\end{thebibliography}
\end{document}